\documentstyle[12pt,epsf]{article}
\def\lsim{\mathrel{\rlap {\raise.5ex\hbox{$ < $}}
{\lower.5ex\hbox{$\sim$}}}}

\newcommand{\pr}{\paragraph{}}
\newcommand{\be}{\begin{equation}}
\newcommand{\ee}{\end{equation}}
\newcommand{\bea}{\begin{eqnarray}}
\newcommand{\nn}{\nonumber}
\newcommand{\eea}{\end{eqnarray}}
\newcommand{\nd}[1]{/\hspace{-0.6em} #1}
\newcommand{\nk}{\noindent}
\def\gappeq{\mathrel{\rlap {\raise.5ex\hbox{$>$}}
{\lower.5ex\hbox{$\sim$}}}}

\def\lappeq{\mathrel{\rlap{\raise.5ex\hbox{$<$}}
{\lower.5ex\hbox{$\sim$}}}}

\begin{document}

\begin{titlepage}
\begin{flushright}  
ACT-16/97 \\
CTP-TAMU-47/97 \\
OUTP-97-67P \\
\end{flushright}
\begin{centering}
\vspace{.2in}
{\large {\bf Decoherent Scattering of Light Particles in a $D$-Brane
Background }
} \\
\vspace{.4in}
{\bf John Ellis$^{a} $},
{\bf P. Kanti}$^{b}$,
{\bf N.E. Mavromatos$^{c,\diamond}$},
{\bf D.V. Nanopoulos}$^{d}$\\
and \\
{\bf Elizabeth Winstanley$^{c}$}
\\
\vspace{.03in}
\vspace{.4in}
{\bf Abstract} \\
\vspace{.05in}
\end{centering}
{\small  We discuss the scattering of two light particles in a
$D$-brane background. It is known that, if one light particle strikes
the $D$ brane at small impact parameter, quantum recoil effects induce
entanglement entropy in both the excited $D$ brane and the scattered
particle. In this paper we compute the asymptotic `out' state of a second
light
particle scattering off the $D$ brane at large impact parameter, showing
that it also becomes mixed as a consequence of quantum $D$-brane recoil
effects. We interpret this as a non-factorizing
contribution to the superscattering 
operator $\nd{S}$ for the two light
particles in a Liouville $D$-brane background, that appears 
when quantum $D$-brane
excitations are taken into account.}
\vspace{0.4in}

\nk $^a$ Theory Division, CERN, CH-1211, Geneva, Switzerland,  \\
$^{b}$ Division of Theoretical Physics, Physics Department,
University of Ioannina, Ioannina GR-451 10, Greece, \\
$^c $
University of Oxford, Dept. of Physics
(Theoretical Physics),
1 Keble Road, Oxford OX1 3NP, United Kingdom,   \\
$^{d}$ Academy of Athens, Chair of Theoretical Physics, 
Division of Natural Sciences, 28 Panepistimiou Ave., Athens GR-10679,
Greece, Center for
Theoretical Physics, Dept. of Physics,
Texas A \& M University, College Station, TX 77843-4242, USA,
and Astroparticle Physics Group, Houston
Advanced Research Center (HARC), The Mitchell Campus,
Woodlands, TX 77381, USA. \\
$^{\diamond}$ P.P.A.R.C. Advanced Fellow.

\vspace{0.01in}
\begin{flushleft}
ACT-16/97 \\
CTP-TAMU-47/97 \\
OUTP-97--67P \\
November  1997\\
\end{flushleft}
\end{titlepage}
\newpage

\section{Introduction and Summary}
\pr
The fate of quantum coherence in the presence of topologically
non-trivial space-time backgrounds is currently the subject of
intensive research. The problem may be posed at two levels: that
of quantum fields in semi-classical gravitational backgrounds
such as macroscopic black holes, and in the presence of
non-perturbative
microscopic quantum fluctuations in the space-time background.
At the semi-classical level, the pioneering work of Hawking and Bekenstein
demonstrated that a description in terms of pure quantum-mechanical
states could not be maintained. Proceeding to the microscopic
quantum level, Hawking~\cite{Hawking} has argued that the conventional
rules of
quantum field theory require some revision, and that particle scattering
should be described by a scattering operator $\nd{S}$ that does not
factorize as a product of $S$ and $S^{\dagger}$ matrix elements: 
\be
\rho_{out} = \nd{S} \rho_{in}: \,\,\, \nd{S} \ne S S^{\dagger}
\label{hawk}
\ee
If this is
the case, there should be a modification~\cite{EHNS} of the conventional
Liouville equation for the time evolution of the density matrix $\rho$:
\be
\partial _t \rho =  i [\rho, H] + \nd{\delta H}\rho
\label{ehns}
\ee
where $H$ is the conventional Hamiltonian. This formulation is
reminiscent of open systems, with Planck-scale degrees of freedom
playing the r\^ole of the unobserved `environment' through which
light low-energy particles move.
\pr
This possibility has been analyzed from various points of
view in recent years. In particular, it has been argued that
behaviours of the form (\ref{hawk}) and (\ref{ehns}) are
inevitable in the non-critical Liouville string approach,
with the non-factorizability of $\nd{S}$ and the magnitude
of $\nd{\delta H}$ related explicitly to departures from the
criticality that defines classical string vacua~\cite{EMN}. Pursuing a
more conventional quantum-gravity approach, divergences in
the partition function of a scalar field in a semi-classical
Einstein-Yang-Mills background have been exhibited, which
have been interpreted~\cite{EMNW} as leading to a modified quantum
Liouville equation of the type (\ref{ehns}).
\pr
Recent advances in $D$-brane
technology~\cite{Dbrane} have provided more powerful tools for analyzing
these issues. In particular, it has been shown that $D$ branes
provide {\it in principle} an exact accounting for the quantum states of black
holes~\cite{counting}.
On the other hand, it has been argued that not all these states are
{\it in practice} measurable in feasible low-energy experiments, and
specific calculations~\cite{EMND} exhibit a loss of information and 
corresponding gain in entropy in the scattering of a light
low-energy particle off a massive $D$ brane, associated with
quantum recoil effects.
\pr
The purpose of this paper is to link these different
approaches to the quantum coherence problem. We first build
on the work of~\cite{EMND} by computing the Riemann, Ricci 
and curvature tensors associated with the singular metric
that describes quantum $D$-brane recoil. This metric had previously
been derived using conformal field theory techniques, and has
many of the properties argued for independently in~\cite{HR}.
We then proceed to compute the effective action that
characterizes long-wavelength physics in the presence
of this $D$-brane recoil. We find
singularities associated with the 
quantum recoil of the $D$ brane induced by
the impact of the first scalar particle, 
and compute properties
of a low-energy scalar field in the presence of such a
recoiling $D$ brane, exhibiting a non-trivial Bogolubov
transformation and particle production. These effects are
sensitive to quantum fluctuations in the $D$-brane recoil, 
which may be treated as a stochastic source. This induces
a non-trivial influence functional, which in turn makes a non-trivial
contribution to the evolution of the density matrix for the external light
particle, as foreseen in
(\ref{hawk},\ref{ehns})~\cite{EHNS,EMN}~\footnote{The quantum origin of
this effect is manifest in the associated annulus topology of
the corresponding world sheet, which in turn corresponds to a loop
diagram in field theory, as also argued in~\cite{HR}. There is no
decoherence at tree level, as also argued in~\cite{Amati}.}.
This
calculation may be regarded as a contribution to low-energy scalar-scalar
scattering in a $D$-brane black-hole 
background, incorporating
the effects of quantum $D$-brane recoil, whose magnitude we estimate. 
Our approach opens the way towards a representation of the
associated  non-factorizing terms in the $\nd{S}$ matrix description
as a spectral integral over $D$-brane states
weighted by their recoil properties.

\section{Space-Time Curvature Induced by Quantum $D$-Brane Recoil}
\pr
We consider the collision of two low-energy, light closed-string
particles in a $D$-brane background, examining the possibility
of a loss of quantum coherence that cannot be accommodated in a
conventional $S$-matrix description. We consider explicitly a
configuration in which one closed-string state impacts the
$D$ brane and has a `hard' scattering off it, while the second light state
passes by at large impact parameter. We have shown previously
that the colliding particle loses coherence as a result of the
necessary sum over $D$-brane states excited by the collision~\cite{EMND}.
Our interest here is whether the more distant light particle
also loses some coherence during the scattering process.

\pr
A necessary preamble to this discussion of light-particle
scattering in a $D$-brane background is a brief review of
relevant aspects of our previous analysis of quantum effects
in the recoil of the $D$ brane struck by the first low-energy light
particle~\cite{EMND}. We have demonstrated that this quantum recoil
problem is described in conformal field theory language by
a pair of logarithmic operators~\cite{recoiltime} $C,D$, which are conjugate
to the $D$-brane position $y_i$ and velocity $u_i$. The
`kick' provided by the incident low-energy particle requires
the introduction of a Heaviside step function, for which we
adopt the following integral representation:
\begin{equation}
\Theta_{\epsilon} (t) = - i \int^{\infty}_{-\infty}
{dq \over q - i \epsilon} e^{i q t} : \epsilon \rightarrow 0^+
\label{theta}
\end{equation}
where $\epsilon$ is an infrared regulator parameter. The
quantum treatment of $D$-brane recoil necessitates the
introduction of world-sheet annulus diagrams, whose
large-size limit is characterized by an infrared regulator
size parameter $L$ that, together with a conventional
ultraviolet world-sheet regulator parameter $a$, specifies
the value of $\epsilon$~\cite{recoiltime}:
\begin{equation}
{1 \over \epsilon^2} \simeq 2 \hbox{log} |L/a|^2
\label{epsilon}
\end{equation}
Further, in our interpretation, we identify the Liouville
field $\phi$ with the renormalization scale on the world 
sheet~\cite{EMN,kogan},
and its zero mode, $\phi_0$, is further identified with the time variable
\begin{equation}
\phi_0 \, = \, t \, \simeq \, \hbox{log} | L / a|
\label{time}
\end{equation}
Note that there is no absolute time in this approach, since physical
quantities are described by the renormalization group,
which relates different scales $L, L'$ that correspond to time
differences $\delta t \simeq \hbox{log}|L/L'|$.

\pr
These identifications of the renormalization scale, the
zero mode of the Liouville field and the time variable
are supported by various consistency
checks, notably of momentum and energy conservation, which
are documented in~\cite{EMND} and references therein.
On the basis of this analysis, we derived the following
form for the singular part of the target-space metric $G_{MN}$
around the moment of the collision:
\begin{equation}
G_{00} = -1, \, G_{ij} = \delta_{ij}, \,G_{0i} = G_{i0} = \epsilon\,
(\epsilon y_i + u_i t) \Theta_{\epsilon} (t) 
\label{metric}
\end{equation}
It is to be understood that, in addition to (\ref{metric}), there
is also a static, spherically-symmetric part of the metric, which
is ${\cal O}(M / R)$ at large distances $R$ from the struck $D$ brane
of mass $M$. We will consider the scattering of a second low-energy
light particle at large impact parameter $R$, so that we may 
neglect this spherically-symmetric part
in a first approximation, and consider the asymptotic
metric as flat to zeroth order in $M / R$. The physics that interests
us is that associated with the $\epsilon$-dependent singularity in
(\ref{metric}).

\pr
We have already pointed out that (\ref{metric}) implies a singularity
in the Riemann curvature scalar as $t \rightarrow 0$~\cite{EMND}. Our
first
priority in this paper is to compute this more explicitly.
To do this, we consider
the following more general form of $D$-dimensional metric:
\begin{equation}
G_{00}=-1 \,,\, G_{ij}=\delta_{ij} \,,\,
G_{0i}=G_{i0}=f_i(y_i,t)\, ,\,\,i,j=1,...,D-1
\label{yiotametric}
\end{equation}
It is easy to check that, for the above metric, the only non-vanishing
components of the fully covariant form of the Riemann curvature
tensor $R_{\mu\nu\rho\sigma}$ are the following:
\begin{eqnarray}
R_{0i0i}&=&\frac{\partial^2 f_i}{\partial y_i \partial t}-
\frac{1}{(1+\sum_{k=1}^{D-1}f_k^2)}\,
\frac{\partial f_i}{\partial y_i}\,
\left(\sum_{k=1}^{D-1}\,f_k\,
\frac{\partial f_k}{\partial t}\right)\\[4mm]
R_{ijij}&=&\frac{1}{(1+\sum_{k=1}^{D-1}f_k^2)}\,
\frac{\partial f_i}{\partial y_i}\,
\frac{\partial f_j}{\partial y_j}
\label{Riemann}
\end{eqnarray}
Correspondingly, the different components of the Ricci tensor
$R_{\mu\nu}$ take the following forms:
\begin{eqnarray}
R_{00}&=& -\frac{1}{(1+\sum_{i=1}^{D-1} f_i^2)^2}\,
\left(\sum_{i=1}^{D-1} f_i \frac{\partial f_i}{\partial t}
\right)\,\left[\sum_{j=1}^{D-1} \frac{\partial f_j}
{\partial y_j} \, \left(1+\sum_{k=1, k\neq j}^{D-1} f_k^2
\right)\right] \\[3mm]
&+& \frac{1}{(1+\sum_{i=1}^{D-1} f_i^2)}\,\left[\sum_{i=1}^{D-1}
\frac{\partial^2 f_i}{\partial y_i \partial t}
\left(1+\sum_{j=1, j\neq i}^{D-1} f_j^2\right)
\right]\\[5mm]
R_{ii}&=&\frac{1}{(1+\sum_{k=1}^{D-1} f_k^2)^2}\,
\left\{\,\frac{\partial f_i}{\partial y_i}\,
\left(\sum_{j=1}^{D-1} f_j\,\frac{\partial f_j}
{\partial t}\right)-(1+\sum_{k=1}^{D-1} f_k^2)\,
\frac{\partial^2 f_i}{\partial y_i \partial t}\right.\nonumber \\[3mm]
&+& \left. \frac{\partial f_i}{\partial y_i} \left[
\sum_{j=1, j\neq i}^{D-1}\, \frac{\partial f_j}
{\partial y_j}\,(1+\sum_{k=1, k\neq j}^{D-1} f_k^2)
\right]\right\} \\[5mm]
R_{0i}&=& \frac{f_i}{(1+\sum_{k=1}^{D-1} f_k^2)^2}\,
\left\{\frac{\partial f_i}{\partial y_i}\,
\left(\sum_{j=1}^{D-1} f_j \,\frac{\partial f_j}
{\partial t} \right)-\left(1+\sum_{k=1}^{D-1} f_k^2\right)\,
\frac{\partial^2 f_i}{\partial y_i \partial t}\right\}\\[5mm]
R_{ij}&=& \frac{1}{(1+\sum_{k=1}^{D-1} f_k^2)^2}\,
f_i \,f_j\,\frac{\partial f_i}{\partial y_i}\,
\frac{\partial f_j}{\partial y_j} 
\label{Ricci}
\end{eqnarray}
\noindent
Finally, the Riemann curvature scalar $R$ is given by
\begin{eqnarray}
R &=& \frac{2}{(1+\sum_{i=1}^{D-1} f_i^2)^2}\,
\left\{- \left( 1+\sum_{i=1}^{D-1} f_i^2 \right) \,\left(
\sum_{i=1}^{D-1} \frac{\partial^2 f_i}{\partial y_i
\partial t} \right)+\left(\sum_{i=1}^{D-1} f_i\,
\frac{\partial f_i}{\partial t}\right)\,\left(
\sum_{i=1}^{D-1} \frac{\partial f_i}{\partial y_i}
\right) \right.  \nonumber \\[3mm]
&+& \left. \sum_{i,j=1, i \neq j}^{D-1} \,
\frac{\partial f_i}{\partial y_i}\,\frac{\partial f_j}
{\partial y_j}\, \left( 1+\sum_{k=1,k \neq i,j}^{D-1} f_k^2
\right) \right\}
\label{curvature}
\end{eqnarray}
It is a trivial matter to insert into the above expressions
the specific form of the metric (\ref{metric}) found in
the previous quantum $D$-brane recoil calculation. We
find the following expressions for curvature scalars
\begin{eqnarray}
R_{\mu\nu\rho\sigma}\,R^{\mu\nu\rho\sigma}&=& 
4\,(D-1)\,\epsilon^4\,[\delta_{\epsilon} (t)]^2 + {\cal O}(\epsilon^6)
\nn \\[3mm]
R_{\mu\nu}\,R^{\mu\nu}&=&
D\,(D-1)\,\epsilon^4\,[\delta_{\epsilon} (t)]^2+ {\cal O}(\epsilon^6)
\nn \\[3mm]
R^2 &=& 4\,(D-1)^2\,\epsilon^4\,[\delta_{\epsilon} (t)]^2
+{\cal O}(\epsilon^6)
\label{repsilon}
\end{eqnarray}
where $\delta_{\epsilon}(t)$ is the appropriate derivative
of $\Theta_{\epsilon}(t)$ (\ref{theta}),
that will be used in the next section.
The reader should not be alarmed by the appearance of the
$[\delta_{\epsilon}(t)]^2$ factors, which do not make 
$\epsilon$-dependent divergent
contributions to the physical quantities of interest to us, such as the
integrated central charge
deficit $Q$, etc., even in the limit $\epsilon \rightarrow 0^+$.
In our regularization, it can be easily shown that in this
limit $\Theta_{\epsilon}(0) \rightarrow \pi$ whilst
$\delta_{\epsilon}(0)$ becomes formally a linearly-divergent
integral that is independent of $\epsilon$.

\pr
We note that the derivation of the space-time discussed above was
essentially non-relativistic, since we worked in the approximation 
of a very heavy $D$ brane. This is the reason why the singularity
in the geometry (\ref{curvature}) is space-like, and why
the consequent change in the quantum state also occurs on a
space-like surface.
This apparent non-causality is merely an artifact of our approximation.
We expect that, in a fully relativistic $D$-brane approach,
the space-time singularity would travel along a light-cone, and that
the quantum state would also change causally.  
Here we are interested in the difference between early- and late-time
quantum states, and this apparent non-causal behaviour 
is irrelevant.

\section{Derivation from an Effective Action}
\pr
We expect that the metric (\ref{yiotametric}) can be derived
using the equations of motion for a suitable singular
effective action induced by quantum $D$-brane effects.
The most naive guess for a possible form of the action is
\begin{equation}
S=\int d^Dx \sqrt{-G}\,\alpha' e^{-2\phi}\,\hat{R}_{GB}^2
\label{action}
\end{equation}
where $\phi$ is the dilaton field and $\hat{R}_{GB}^2$ is
the ghost-free Gauss-Bonnet quadratic combination of the
Riemann tensor, Ricci tensor and curvature scalar:
\begin{equation}
\hat{R}_{GB}^2 = R_{\mu\nu\rho\sigma}\,
R^{\mu\nu\rho\sigma}-4 R_{\mu\nu}\,R^{\mu\nu}+R^2
\label{gaussbonnet}
\end{equation}
This is indeed the case, as we now show.

\pr
The field equations derived from the action 
(\ref{action}) and (\ref{gaussbonnet}) are
\begin{eqnarray}
\alpha'\,\hat{R}_{GB}^2&=& \alpha' (R_{\mu\nu\rho\sigma}\,
R^{\mu\nu\rho\sigma}-4 R_{\mu\nu}\,R^{\mu\nu}+R^2)=0 \\[4mm]
\alpha' K_{\mu\nu}&\equiv&\alpha'\,(R\,\phi_{;\,\mu\nu}-
R\,\phi_{;\,\,\rho}^{\,\,\rho}\,G_{\mu\nu}
-2\,R_{\mu\rho}\,\phi_{;\,\nu}^{\,\,\,\,\,\rho}-2\,
R_{\nu\rho}\,\phi_{;\,\mu}^{\,\,\,\,\,\rho} \nonumber \\[3mm]
&+&2\,R^{\rho\sigma}\,\phi_{;\,\rho\sigma}\,G_{\mu\nu}+
2\,R_{\mu\nu}\,\phi_{;\,\,\rho}^{\,\rho}
-2\,R_{\mu\rho\nu\sigma}\,\phi_{;}^{\,\rho\sigma})
\nonumber \\[3mm]
&-&2\alpha'\,(R\,\phi_{;\,\mu}\phi_{;\,\nu}-
R\,\phi_{;}^{\,\rho}\,\phi_{;\,\rho}\,G_{\mu\nu}-
2\,R_{\mu\rho}\,\phi_{;\,\nu}\,\phi_{;}^{\,\rho}-
2\,R_{\nu\rho}\,\phi_{;\,\mu}\,\phi_{;}^{\,\rho} \nonumber \\[3mm]
&+&2\,R^{\rho\sigma}\,\phi_{;\,\rho}\,\phi_{;\,\sigma}\,G_{\mu\nu}+
2\,R_{\mu\nu}\,\phi_{;}^{\,\rho}\,\phi_{;\,\rho}-
2\,R_{\mu\rho\nu\sigma}\,\phi_{;}^{\,\rho}\,
\phi_{;}^{\,\sigma})=0
\label{motion}
\end{eqnarray}
where $;$ denotes covariant differentiation. 
In the case of the first equation of motion, we see
immediately from the expressions (\ref{repsilon}) that
\begin{equation}
R_{\mu\nu\rho\sigma}\,R^{\mu\nu\rho\sigma}-
4\,R_{\mu\nu}\,R^{\mu\nu}+ R^2={\cal O}(\epsilon^6)
\label{yiotagaussbonnet}
\end{equation}
Therefore, the first equation of motion (\ref{motion})
is satisfied to leading order in the small parameter $\epsilon$.

\pr
Motivated by the Liouville string theory approach of ref. \cite{EMND},
we use the linear-dilaton configuration~\cite{aben,EMN}
\be
 \phi = Qt 
\label{lineardilaton}
\ee
where $Q^2$ is the central-charge deficit of the Liouville string, and 
$Q = {\cal O}(\epsilon^2)$, as we shall argue below. 
This linear 
dependence of the dilaton field on the time $t$ is expected to be valid 
approximately at large times $t \sim 1/\epsilon^2 >> 0$ after the
collision, 
where the system is close to its non-trivial world-sheet 
fixed point~\cite{EMND,EMN}. This is sufficient for our purposes
to demonstrate particle production and decohering effects,
as we shall argue in the next sections. 
Using (\ref{lineardilaton}), we obtain for the components of the
second (Einstein) equation of motion (\ref{motion}):
\begin{eqnarray}
K_{00}&=& {\cal O}(\epsilon^8) \\[3mm]
K_{0i}&=& 2\,Q\,f_i\sum_{j,k=1,\,j\neq k \neq i}^{D-1}
\frac{\partial f_j}{\partial y_j}\,
\frac{\partial^2 f_k}{\partial y_k\,\partial t} +
{\cal O}(\epsilon^8) \nonumber \\[3mm]
&=& 2\,(D-2)\,(D-3)\,Q\,\epsilon^5\,u_i\,t\,
\Theta_{\epsilon}^2(t)\,\delta_{\epsilon}(t)+ {\cal O}(\epsilon^8) \\[3mm]
K_{ii}&=& 2\,Q \sum_{j,k=1,\,j\neq k \neq i}^{D-1}
\frac{\partial f_j}{\partial y_j}\,
\frac{\partial^2 f_k}{\partial y_k\,\partial t} +
{\cal O}(\epsilon^8) \nonumber \\[3mm]
&=&
2\,(D-2)\,(D-3)\,Q\,\epsilon^4\,\Theta_{\epsilon}(t)\,\delta_{\epsilon}(t)
+ {\cal O}(\epsilon^8) \\[3mm]
K_{ij}&=& {\cal O}(\epsilon^{10})
\end{eqnarray}
Thus, we again conclude that the equations of 
motion are satisfied at least through ${\cal O}(\epsilon^6)$.

\pr
To see that the central charge deficit $Q={\cal O}(\epsilon^2)$, we
proceed as follows. We recall that $Q^2$ is the 
change in the Zamolodchikov $C$ function over the  entire 
world-sheet renormalization-group trajectory from the 
Gaussian to the non-trivial fixed point. 
In our interpretation of target time as 
the world-sheet Liouville scale, this results for our purposes
in the following space-time integrated expression:
\be
     Q^2 \sim \int dt \beta^i {\cal G}_{ij} \beta ^j 
\label{Cfunction}
\ee
where $\beta ^i $ is the $\beta$ function of 
a generic string background characterized by couplings
$g^i$, and ${\cal G}_{ij}$ is the corresponding Zamolodchikov metric. 
In our case, 
to lowest non-trivial order  in $\alpha '$, the only background that 
contributes to ${\cal G}$ is the target metric $G_{ij}$~\footnote{The
dilaton $\beta$ function 
is already ${\cal O}(\alpha ')$ compared to the graviton.},
and the corresponding Zamolodchikov metric is given by
\be 
    {\cal G}_{\mu\nu\alpha\beta} = a [G_{\mu\nu} G_{\alpha\beta} 
- G_{\mu\alpha}G_{\beta\nu} - G_{\mu\beta}G_{\alpha\nu} ] + {\cal O}(\alpha ') 
\label{zamimetric}
\ee
where $a$ is an arbitrary constant~\cite{mm}. 
The lowest-order graviton $\sigma$-model $\beta$-function 
is:
\be
   \beta_{\mu\nu}^G=\alpha '\{ R_{\mu\nu} + 2 \nabla _\mu \nabla _\nu
\Phi  \} + {\cal O}[(\alpha ')^2] 
\label{beta}
\ee
so that
the integrand in (\ref{Cfunction}) 
simplifies to
\be
\beta^i {\cal G}_{ij} \beta ^j =(\alpha ')^2 a 
\epsilon ^4 \left[2(D-1)(D-2)[\delta_{\epsilon} (t)]^2\right] + {\cal
O}(\epsilon
^6) .
\label{integrand}
\ee
to leading order in $\epsilon$.

\pr
The above expression
(\ref{integrand}) 
is understood to be multiplied 
by an overall normalization factor of the form $W=\left[1 + \sum_{i=1}^{D-1}
\epsilon ^2 \left(\epsilon y_i + u_i t\right)^2 \Theta^2 (t) \right]^{-1}$,
which ensures the finiteness of all spatial integrals.
At leading order in $\epsilon$, this factor may be approximated by 
unity. The temporal integration over $t$
yields a divergent constant normalization factor $\delta (0)$, which
is irrelevant because $a$ is arbitrary~\cite{mm}. We conclude that
\be
Q^2\sim 
(\alpha ')^2 a 
\epsilon ^4 \left[2(D-1)(D-2)\delta_{\epsilon} (0)\right] + {\cal
O}(\epsilon ^6) ,
\ee
which implies that $Q$ is of order $\epsilon ^2$ for $D>2$,
as assumed above.

\section{Particle Creation in the $D$-Brane Background}

\pr
In this section we show that an observer who scatters a scalar 
field off the space-time whose geometry is described by
(\ref{metric}) sees particle creation in the singular background
metric of the struck $D$ brane. 
To this end, we first note that an on-mass-shell
scalar field $\phi $ of mass $\mu$ in the background (\ref{metric})
may be expanded in terms of ordinary flat-space Minkowski modes which
play the r\^ole of the ``in'' modes, since
our background space (\ref{metric}) can be
mapped, for $t > 0$, to a flat space to ${\cal O}(\epsilon ^2)$
by means of a simple coordinate transformation:
\be 
{\tilde t} = t \qquad, \qquad {\tilde X}^i = X^i + \frac{1}{2}\epsilon  u_i t^2 
\label{rindler}
\ee
which may be represented by the Penrose diagram shown
in Fig. 1.

\pr
Each of the four diamonds corresponds in a conventional
Rindler space to one of the causally-separated regions~\cite{birrell}.
Our space-time corresponds to the right-hand diamond.
We see that it is flat
Minkowski for $t < 0$, and that for $t >> 0$ 
the shaded region of the space-time resembles, to ${\cal O}(\epsilon )$,
a Rindler wedge, with `acceleration'  $\epsilon u_i$. The event horizons
are indicated by the straight lines separating the diamond-shaped 
regions of the diagram. The dashed line corresponds to the 
curvature singularity $\epsilon ^2 \delta_{\epsilon} (t)$, which may be
ignored in order ${\cal O}(\epsilon)$.   
As a consequence of the non-relativistic nature of the $D$ brane, 
this singularity  appears to violate causality, lying outside the 
light cone. This is merely an artifact of taking the
limit as the velocity of light 
$c \rightarrow \infty$, as is appropriate for a non-relativistic (heavy) 
$D$ particle. One expects causality to be restored, with a rotation of the 
singular surface so as to become light-like, in a fully relativistic 
treatment of the problem, which lies beyond the scope of the present 
work.

\begin{centering}
\begin{figure}[htb]
\epsfxsize=5in
\centerline{\epsffile{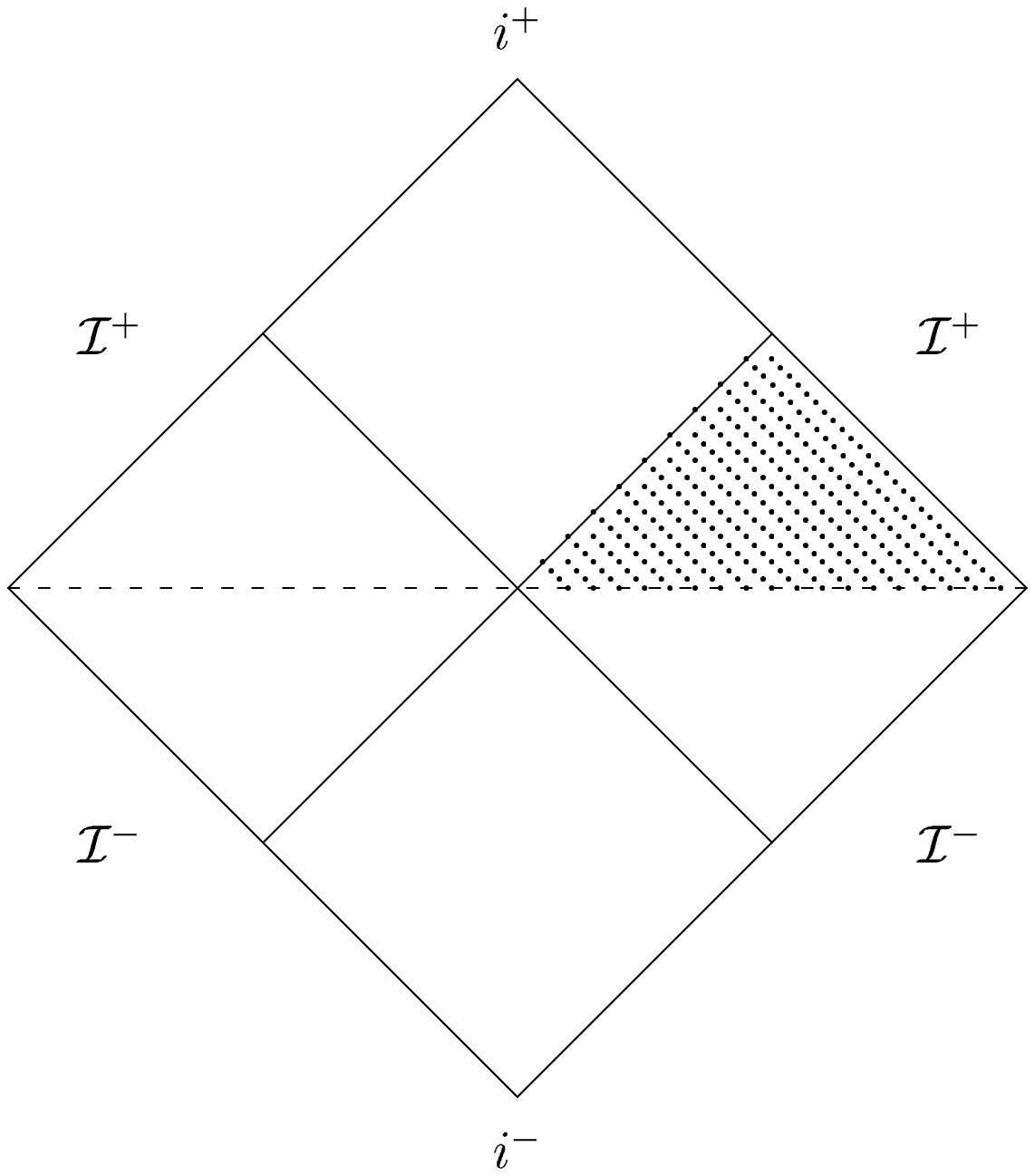}}
%
%
\caption{{\it Penrose diagram for the 
space-time environment derived from the
quantum recoil of a heavy $D$ brane induced by the
scattering of a light closed-string state.}}
\label{fig1}
\end{figure}
\end{centering}

\pr

In view of this figure, we expect~\cite{birrell} to find a non-trivial
Bogolubov transformation between the ``in" and ``out" vacua
at $t < 0$ and $t >> 0$, which we now exhibit.
The Minkowski mode expansion for the scalar field 
takes the standard form: 
\be
 \phi = \int dE d^{D-1} K [a_{EK} {\hat u}_{EK} + a^\dagger_{EK}
{\hat u}^*_{EK} ]
\label{minkowskimodes}
\ee
where
\be
{\hat u}_{EK}=\exp (-iEt+ i{\underline {K}}.{\underline {X}})
\qquad E^2 - {\underline K}^2=\mu^2  
\ee
are the usual Minkowski ``in'' modes. 
Such an expansion is valid for $t < 0$. 
For $t >> 0$ one may expand the
scalar field in terms of ``out'' modes, 
\be
   \phi = \int dE d^{D-1}K [b_{E,K} {\tilde u}_{E,K} 
+ b^\dagger _{E,K} {\tilde u}^*_{E,K}] 
\label{outmodes}
\ee
with 
\be 
{\tilde u}_{EK}=\exp (-iEt + i {\underline K}.{\underline X} + 
i\frac{1}{2}\epsilon {\underline u}.{\underline K} t^2 ) 
+ {\cal O}(\epsilon ^2) 
\label{out}
\ee
These modes
are
related 
to the ``in'' ones by the Bogolubov coefficients $\alpha, \beta$:  
\be 
{\tilde u}_{E,K} = \int dE' d^{D-1}K' [\alpha_{EE'KK'} {\hat u}_{E'K'} + 
\beta_{EE'KK'}{\hat u}^*_{E'K'}]
\label{rindlermodes} 
\ee
Extending appropriately the definitions 
of the ``in'' and ``out'' modes to the entire space-time, 
we find that the Bogolubov coefficients are given by: 
\bea
\alpha_{EE'KK'}= &~&
\int dt d^{D-1}X {\tilde u}_{E,K} {\hat u}^{*}_{E'K'} = \nn \\
&~&\delta^{(D-1)}(K-K')[\delta (E-E') + 
\frac{i}{2}\epsilon ({\underline u}.{\underline K})
\int _0^\infty dt e^{-i(E-E')t}
t^2 + {\cal O}(\epsilon ^2) 
\nn \\
\beta_{EE'KK'}= &~& \int dt d^{D-1}X {\tilde u}_{E,K} {\hat u}_{E'K'} = \nn \\
&~&\delta^{(D-1)}(K+K')\frac{i}{2}\epsilon ({\underline u}.{\underline K})
\int _0^\infty dt e^{-i(E+E')t}t^2 
+ {\cal O}(\epsilon ^2)  
\label{bogol2}
\eea
which satisfy $|\alpha |^2 - |\beta |^2 = 1$ in a distribution-theoretic
sense. The $t$ integrals are divergent, but they are viewed as 
irrelevant constants.   

\pr
Bearing in mind the analogy with Rindler space, we expect
particle creation, which is controlled by
the Bogolubov $\beta$ coefficient,
since it is this term that contributes to positive 
frequency mixing~\cite{birrell}. Indeed, we find that
the number of particles created in the mode labelled by $E$ and $K$
is given by: 
\bea
n_{EK} &~& = ~_{IN}\langle 0|b^\dagger_{EK}b_{EK}|0 \rangle _{IN} =
|\beta _{EKEK}^2| \nn \\
&~& \simeq \left(\epsilon  
{\underline u}.{\underline K}\right)^2  
|\left(\int _0^\infty dt e^{-i2Et}t^2 \right)|^2 \nn \\
&~& \propto \left(\frac{\epsilon ^2 ({\underline u}.{\underline
K})^2}{E^6}  \right)
\label{creation}
\eea
The formula (\ref{creation}) 
describes particle creation with a non-trivial angular
distribution around the direction of the velocity vector
$u_i$. The particle spectrum is {\it not thermal} as it
would have been in the case of a uniform acceleration~\cite{unruh}.

\pr
This particle creation (\ref{creation})
may be interpreted as non-thermal Hawking radiation 
from the recoiling D-brane. An analogous effect was suggested in the
context of non-critical string theory~\cite{EMN}
on the basis of an analysis of two-dimensional 
stringy black holes~\cite{emnrad}, the latter 
being viewed as `massive' string 
particles. It was argued in~\cite{emnrad}  that the
scattering of light matter off the 
black hole excites the black hole to a 
higher string level. Quantum instabilities would then
cause the decay of this excited state, with the inevitable emission 
of non-thermal radiation. 
In a similar context, we mention a related 
work~\cite{diamand}, where the presence 
of non-thermal radiation has been 
demonstrated when a massive scalar field is quantized in a 
two-dimensional dilaton-gravity black-hole background. 
In a large-mass $1/m$ expansion for the scalar field,  
such non-thermal radiation effects appear, associated with 
$R^2/m^2$-corrected terms in the target-space 
effective action. These describe effects due to 
the back reaction of 
the massive matter on the two-dimensional space-time geometry,
and are very analogous to
our higher-dimensional case, in which we also obtained an
effective Gauss-Bonnet $R^2$ target-space action term
(\ref{action}),(\ref{gaussbonnet}),
describing the leading effects of the 
back reaction of the heavy recoiling $D$ brane on the surrounding 
space-time geometry. 

\section{Quantum Background Fluctuations}
\pr
Despite this particle production, we
still need to demonstrate the appearance of entropy,
and thus decoherence, in the
spectator-particle system. According to our interpretation of
the entropy generation as being associated with quantum
effects in $D$-brane recoil, any such
decoherence effects should be related to the quantum excitation
of the $D$ brane, and the resulting sum over internal states. This is
not apparent from the above discussion, which relates the
spectator particle number to the
velocity $u_i$. Particle production could in principle occur in a coherent
way, so how do we know that is not the case here?
The answer is that the quantum summation over
world-sheet genera results in fluctuations
of the velocity $u_i$ about the classical value 
which would be expected on the basis of
momentum conservation in an elastic collision~\cite{lizzi,EMND}.

\pr
Using generic 
arguments~\cite{lizzi} based on the quantization of $\sigma$-model
backgrounds in string theory associated with
the summation over genera~\cite{EMN}, it has been shown that 
the $D$-brane velocity fluctuates about $u_i$ with
a Gaussian distribution
\be
   {\cal P} = e^{-\frac{(\delta u_i)^2}{\Gamma _u^2}}
\qquad \Gamma _u ^2 \sim (g_s^2 + \delta g_s^{2})~{\rm ln}\delta, ~~~\delta \rightarrow 0
\label{velocitydistr}
\ee
where $\delta g_s^2$ is the change 
in the square of the coupling constant for the string, $g_s^2$, 
induced by the matter deformation of the $\sigma$ model describing
the $D$ brane that is caused by the scattering of the
first string particle. The appearance of 
a ${\rm ln}\delta$ dependence is due to world-sheet divergences,
arising from world-sheet loops down to a size $\delta << 1$, which were
argued in~\cite{lizzi} to dominate
the description of velocity fluctuations around a recoiling 
heavy-$D$-brane background.

\pr
It is this divergence related to a pair of 
logarithmic operators~\cite{recoiltime}  
that
explains in conformal-field-theory language why decoherence appears in
scattering off a $D$ brane, but would not appear in scattering off a
conventional massive string state. Physically, the dispersion in recoil
velocity $u_i$ implies that at large times the $D$-brane state must be
described by an ensemble of widely-separated configurations. In addition
to the mixed nature of the $D$-brane state itself, this
distributed ensemble yields a superposition of macroscopically-distinct
space-time backgrounds, leading to entropy growth also for the second
light string state in the scattering process, as we discuss below.

\pr
The change $\delta g_s$ in $g_s$ due to the matter deformation is given 
in conventional $\sigma$-model perturbation theory by: 
\be 
  \delta g_s^2 =e^{-2<\Phi >_g} - e^{-2<\Phi >_0} \simeq 
g_s^2 \sum _{N} <\Phi V_{i_1} \dots V_{i_N} >q^{i_1} \dots q^{i_N} 
\label{change} 
\ee
where the $q^i$ denote background 
target-space couplings
corresponding to vertex operators $V_i$, 
and $<\dots >_g$ ($<\dots >_0)$ 
denotes $\sigma$-model correlators in the deformed (free) string. 
In our case, 
only correlators with even powers of $D$ are non-vanishing, and
an order-of-magnitude estimate of $\delta g_s$ can 
be made using the two-point correlator of the  
recoil operator $D_i$ on the disc, which
diverges as $1/\epsilon $:
\be 
\delta g_s^2 =g_s^2 
<\Phi \int \int _{\partial \Sigma} D_i D_j > u^iu^j 
+ \dots \sim 
g_s^2 \frac{1}{\epsilon^2} \times {\cal O}(E^2/M_D^2)
\label{deltags1}
\ee
where $\partial \Sigma$ denotes a world-sheet boundary, 
and 
$E$ is a typical energy
of the light-string state that scatters off the 
$D$-brane background. 
Although 
the expression (\ref{deltags1}) appears formally 
divergent, it is in fact subleading for 
small $u_i << 1$, compared to the $g_s^2$ terms in 
the width (\ref{velocitydistr}), as we discuss later on. 
For now we note that 
higher-order correlators of $D$ may be ignored in the
dilute-gas approximation, which is valid for weak string 
couplings $g_s << 1$. Using a generalized 
Fischler-Susskind mechanism~\cite{fischler}, we may identify
$1 / \epsilon^2 ={\rm ln}\delta$, thereby implying 
the following expression for the dominant terms in the width 
(\ref{velocitydistr}):
\be
\Gamma _u^2 \equiv   (\Delta u)^2 \sim g_s^2{\rm ln} \delta   
+ g_s^2 u_i^2 {\rm ln}^2 \delta    
\label{spread2}
\ee
It is clear from this discussion that
if the string is very weakly coupled: $g_s \rightarrow 0$, 
the quantum fluctuations $\Delta u_i$  will be suppressed,
as is appropriate for a semi-classical heavy $D$ brane of mass 
$M \sim 1/g_s$. 

\pr
As discussed above, here and in~\cite{EMND} we further
identify $1 / \epsilon^2$, and hence $\rm{ln}\delta$, with 
the target time
$t$~\footnote{For an alternative interpretation, see~\cite{lizzi}.}. 
Within this interpretation, we may interpret
(\ref{spread2}) as the spread 
in time $t$ of the probability distribution $|\Psi ^{st}(u_i,t)|^2$
for the string wave function.
This interpretation of the summation 
over topologies in a $\sigma$-model path integral 
has previously been made in~\cite{aspects,emninfl},
where it was shown that such probability distributions 
satisfy stochastic Fokker-Planck diffusion equations. 

\pr
We now show how this effect
leads to stochastic dynamics and decoherence
for the spectator scalar field propagating in the 
background of a $D$ brane with such a quantum-mechanical treatment
of recoil. 
Using the initial coordinate system $(y_i, t)$ 
and the metric (\ref{metric}), we find the following
equation of motion for a massive scalar field $\varphi$, of mass $\mu$,
\bea
&~&-\partial_t^2 \varphi + \sum_{i=1}^{D-1} \left(\partial_i^2 
+ 2f_i\partial_i \partial_t \right)\varphi
+ \sum_{i=1}^{D-1}[\sum_{i \ne j} f_j^2 ] \partial_i^2 \varphi 
- \nn \\
&~& \sum_{i \ne j} f_if_j\partial_i \partial_j \varphi + \rho \partial_t 
\varphi + \sum_{i=1}^{D-1} B_i \partial_i \varphi - \mu^2 W \varphi =0
\label{equations}
\eea
where $f_i  = \epsilon (\epsilon y_i + u_it)\Theta_\epsilon (t)$
and $W = 1 + \sum_{i=1}^{D-1} \epsilon^2 
\Theta_\epsilon (t)^2 (\epsilon y_i + u_i t)^2$ as before,
$\sqrt{-G}=\sqrt{W}$, and
\bea
\rho &=&W^{-1}\left\{\sum_{i=1}^{D-1}\left(
\epsilon^2(\epsilon y_i + u_i t)\Theta _\epsilon (t)\, 
[u_i\Theta _\epsilon (t) 
+ \delta _\epsilon (t)\,(\epsilon y_i + u_i t)\,]\right)
\right. \nn \\[1mm] &+& \left.
(D-2)\epsilon ^2 \Theta _\epsilon (t) W 
+ \epsilon^2 \Theta _\epsilon (t))\right\}, \nn \\[2mm]
B_i&=&\epsilon u_i \Theta _\epsilon (t) + \epsilon (\epsilon y_i + u_i t)
W^{-1}\delta_\epsilon (t) -
f_i \Theta_\epsilon(t) W^{-1}\sum_{j=1}^{D-1}\epsilon u_j f_j \nn \\[1mm]
&-&(D-2) f_i \epsilon ^2 \Theta _\epsilon (t) - 
\epsilon ^2 W^{-1} f_i \Theta _\epsilon (t) 
\label{definitions}
\eea
We see that (\ref{equations}) contains a
friction term, proportional to the quantity $\rho ={\cal O}(\epsilon
^2)$ in (\ref{definitions}),
which is of the same order as the curvature in our space time,
and depends on the $\delta _\epsilon (t)$ singularity structure. 
\pr
To illustrate our basic argument for the stochastic nature 
of the problem, we consider the simplest possible configuration 
for the massive scalar field $\varphi$, namely that of a 
field that is constant in space: $\varphi (t)$. 
In this case (\ref{equations}) simplifies
to:
\be
   -\partial_t^2 \varphi (t) + \rho \partial_t \varphi =\mu^2W \varphi 
\label{inflation1}
\ee
Consistency with the approximation of spatial independence can be
maintained only up to order $\epsilon ^2$,
where the $y_i$ dependent terms in $\rho$ are ignored, and
the latter may be approximated by 
\be
   \rho \simeq \epsilon ^2 u_i^2 t \Theta_\epsilon (t) (\Theta_\epsilon (t) 
+ t\delta _\epsilon (t)) 
+ (D-1)\epsilon ^2 \Theta_{\epsilon} (t) + {\cal
O}(\epsilon ^3)
\label{rho}
\ee
For large times $t \sim 1/\epsilon ^2$, 
the equation of motion (\ref{inflation1}) 
resembles that for an inflaton field
to the order in $\epsilon$ studied, with `friction' $\sim u_i^2$, 
and a time-dependent force
that increases linearly with large times $t >> 0$: 
\be
   -\partial_t^2 \varphi (t) + u_i^2  
\Theta_\epsilon (t)\partial_t \varphi =\mu^2 
\varphi  - \mu^2 u_i^{2} t \Theta_{\epsilon} (t)\varphi 
\label{inflation}
\ee
Thus there are
two possible sources of decoherence in our problem. One 
is related to the friction term in (\ref{inflation}), 
which for any finite time $t < 1/\epsilon ^2$, is of the same
order in $\epsilon $ as the space-time curvature, i.e., of ${\cal
O}(\epsilon ^2)$, and hence can be ignored
in our present discussion. The other is due to
the interaction of the 
scalar field $\varphi$ with the gravitational background - recall that $W$
is the determinant of the metric tensor - and therefore
provides~\cite{emninfl} an environmentally-induced stochastic source term
in the inflation equation (\ref{inflation}), 
associated with the quantum recoil degrees
of freedom of the $D$ brane,
which survives to order $\epsilon$ 
in the flat space-time limit that
we are considering here. It is this source that 
we now explore further, and use to estimate the size of the 
induced decoherence effects. 
\pr
A general approach to the analysis of such a system,
described by field
variables $a$ in interaction 
with some environmental degrees of freedom $q$
with some action:
$  S[a,q]=S[a] + S_e[q] + S_{int} [a,q] $, is provided in~\cite{vernon,hu}.
The system $a$ is characterized by a
reduced density matrix $\rho_r (a,a')$:
\be
\rho _r (a,a')=\int _{-\infty}^{\infty} dq \int _{-\infty} ^{\infty} 
dq' \rho (a,q; a', q')\delta (q-q') 
\label{density}
\ee
whose time evolution is given by the evolution operator
${\cal J}_r$:
\be
\rho _r (a,a',t)= \int \int_{-\infty}^{\infty} da_i da'_i
{\cal J}_r(a,a',t|a_i,a_i',t_i)\times \rho_r(a_i,a_i',t_i)
\label{evol}
\ee
with the subscript $i$ denoting initial data at a time $t=t_i$,
when the system and the environment are 
assumed to be uncorrelated: ${\rho} (t_i)= {\rho_r}(t_i)
\times {\hat \rho}_{environment}(t_i)$. Using the
path-integral formalism~\cite{vernon}, it has been shown that
the evolution operator can be expressed in terms of the influence
functional ${\cal F}(a,a')$:
\be
{\cal J}_r (a_f,a_f',t|a_i,a'_i,t_i)=
\int_{a_i}^{a_f} \int _{a_i'}^{a_f'} Da Da' 
e^{\frac{i}{\hbar}\{ S[a] - S[a'] \}}{\cal F}(a,a')
\label{infulence}
\ee
with the subscript $f$ denoting the `final' state at some large time $t$.
A formal representation of the 
influence functional is given by~\cite{vernon,hu}: 
\bea
&~&{\cal F}(a,a') \equiv e^{\frac{i}{\hbar}S_{IF}} = 
\int \int \int _{-\infty}^{\infty}
dq_f dq_i dq'_i \int_{q_i}^{q_f}Dq \int _{q_i'}^{q_f'}Dq'
{\rm exp}\left(\frac{i}{\hbar}
\{ S_{e}[q] + S_{int}(a,q)\}\right)
\nn \\
&~&{\rm exp}\left(-\frac{i}{\hbar}\{S_{e}[q'] + S_{int}(a,q') \}\right)  
\rho_e(q_i.q_i',t_i)
\label{if}
\eea
There is a simple representation~\footnote{The 
derivation of~\cite{hu}, which was based on some 
cosmological models, can easily be generalized to 
more general cases including the one we consider here,
using a finite-volume regularization.}
of the influence functional in terms of the 
Bogolubov coefficients $\alpha_k(a)$,$\beta_k(a)$ 
that appear in the transformation 
between the creation and annihilation operators 
of field amplitudes at different times:
\be
    {\cal F}(a,a')=\prod _{k} \frac{1}{\sqrt{\alpha _k(a')\alpha^*_k(a)
-\beta_k (a')\beta _k^*(a)}}
\label{bogolinfl}
\ee
Note that since the Bogolubov coefficients in general satisfy 
$|\alpha(a)|^2 - |\beta (a)|^2 =1$, the influence 
functional reduces to the unit matrix when $a'=a$, as it should. 

\pr
In the application of this formalism to our problem, the r\^ole
of the field variables $a$ at an initial time $t_i \le 0$  
is played by the Minkowski ``in'' modes ${\hat u}_{EK}$ of the previous 
section, whilst the r\^ole of the field variables $a'$, at a later time 
$t >>0$, is played by the ``out'' modes
${\tilde u}_{EK}$ (\ref{out}), but corresponding to a velocity 
which is given by the quantum fluctuations (\ref{spread2}). 
We shall consider decoherence 
between two field configurations that preserve energy and momentum. 
This is a feature which is respected on average by our 
Liouville approach to target time~\cite{EMN}. 
 
\pr
{}From the expressions (\ref{bogol2})
for the Bogolubov coefficients
we find the following
order of magnitude of the pertinent 
influence action $S_{IF}$ (\ref{if}),(\ref{bogolinfl}): 
\bea
S_{IF} &~& \propto -i \hbar {\rm ln}{\cal F}(a,a') \sim 
i \hbar [\frac{1}{2}{\rm ln}\left(\alpha _{EKEK}({\tilde u}_{EK})\right) 
+ \dots ] \nn \\
&~& \sim i\hbar 
[\frac{1}{4}g_s^2 |{\underline {\hat u}}.{\underline K}|^2~t  + 
{\cal O}(g_s^2 u_i^2|{\underline {\hat u}}.{\underline K}|^2~t^2)]  
+ \dots 
\label{final}
\eea
where ${\underline {\hat u}}$ 
denotes a unit vector in the direction of ${\underline u}$, and 
the $\dots $ indicate real parts, 
which are not of interest to us here,
as well as terms of higher 
order in $\epsilon ^2$, which are suppressed as $\epsilon  \rightarrow 0$.
We see that
the term in (\ref{final}) proportional to $t^2$, which owes its
existence to the 
$\delta g_s^2$ term in (\ref{velocitydistr}), 
is subleading for small $u^2 << 1$,
compared to the term with a linear $t$ 
dependence. This follows from the fact that,
for asymptotically large times $t \sim 1/\epsilon ^2$, these terms 
are of order $(u_i^2/\epsilon ^2) t $. It has been
argued in~\cite{EMND,lizzi}
that $u_i/\epsilon = u_i^R $ is a `renormalized' velocity, corresponding 
to the $\sigma$-model coupling of an exactly-marginal velocity 
recoil operator $D$, which in our framework is also assumed small 
$u_i ^R << 1$.     

\pr
Thus, we see that
the quantum fluctuations of the $D$-brane recoil velocity (\ref{spread2}) 
induce decoherence for the second light particle that grow linearly in
time, provided that the modular world-sheet 
infinities ${\rm ln}\delta$ have the target-time
interpretation~\cite{EMND} assumed here.  
If only classical recoil velocity 
were present, there would be no ${\rm ln}\delta$ dependence
and hence no time dependence and no decoherence.
The imaginary coefficient in (\ref{final}), and its linear
dependence on $t$, imply that the quantum recoil of the $D$ brane,
which corresponds in conformal-field-theoretic language to a
departure from criticality, induces 
a non-Hamiltonian contribution $\nd{\delta H}$ (\ref{ehns})
into the generic
evolution equation of the reduced density matrix for
the scalar particle in this background. 
Since 
the momentum of the 
spectator particle, $K$, may be taken generically to be of the same order 
of magnitude as  
its energy $E$,  
one obtains from (\ref{final})
the order-of-magnitude
\be 
\nd{\delta H} \sim {\cal O}( E^2/M_D)
\label{hslash}
\ee
where $M_D$ is the heavy $D$-brane mass, which is related to the 
conventional string scale $M_s=(\alpha ')^{-1/2}$, with $\alpha'$ the Regge 
slope, via  
$g_s \sim M_s/M_D < 1$~\footnote{Here we assume that 
the dimensionless time scale $t \sim {\rm ln}\delta $ 
is expressed in units of $M_D^{-1}$. Further analysis
is needed to pin down the most appropriate scaling. If
instead one assumes
that the dimensionless time scale ${\rm ln}\delta$ is expressed in string 
units $M_s^{-1}$, then (\ref{hslash}) has an extra factor $g_s$.}.  
The formula (\ref{hslash}) agrees
with the generic estimates of~\cite{EHNS,EMN,EMND},  
which indicated a suppression by a single power of the heavy mass scale, 
in this case $M_D$. 

\section{Conclusions and Outlook}  

\pr
We had previously shown explicitly using conformal field theory
techniques~\cite{EMND} that a light particle that has a `hard'
scattering off a $D$ brane finishes up in a mixed
state, with an entanglement entropy corresponding to that of the
excited struck $D$ brane. The analysis in the previous section
demonstrates that a light spectator particle passing by at large
impact parameter is also converted into a mixed state,
by virtue of the quantum fluctuations in the recoil of the
struck $D$ brane. This can be
regarded as a decohering contribution to the scattering of
light particles in a $D$-brane background, and hence a 
non-trivial contribution to the $\nd{S}$ matrix, over and above
the decohering effects of $D$-brane scattering on the
propagation of a light particle that we discussed previously~\cite{EMND}.

\pr
More specifically, we have demonstrated that this decoherence
effect causes off-diagonal density matrix elements to 
decay exponentially with time, just as expected in~\cite{EHNS},
on the basis of the non-Hamiltonian term $\nd{\delta} H$ in (\ref{ehns}).
Moreover, this effect increases quadratically with the particle energy
$E$, as also suggested previously~\cite{EHNS,EMN,EMNW}.

\pr
In the future, it would clearly be interesting to extend this
analysis to light-particle scattering in the absence of an
initial-state $D$ brane, but taking into account the excitation of
intermediate virtual $D$-brane states. Such a computation goes beyond the
scope of this paper, but we expect it to share features with this
calculation, in particular that it would lead to a contribution to
$\nd{\delta} H \sim E^2$.
\pr

\pr
\nk {\bf Acknowledgements}
\pr
P.K. would like to thank the CERN Theory Division for its
hospitality and financial support. 
The work of D.V.N. is supported in part by D.O.E. Grant
DE-FG03-95-ER-40917, and that 
of E.W. is supported by Oriel College, Oxford.


\begin{thebibliography}{99}
\bibitem{Hawking} S.W. Hawking,  Comm. Math. Phys. 87 (1982), 395.
\bibitem{EHNS} J. Ellis, J.S. Hagelin, D.V. Nanopoulos and
M. Srednicki, Nucl. Phys. B241 (1984), 381.
\bibitem{EMN} J. Ellis, N.E. Mavromatos
and D.V. Nanopoulos, Phys. Lett. B293 (1992), 37;
\par Mod. Phys. Lett. A10 (1995), 425, and
hep-th/9305117;
\par Lectures presented at the
{\it Erice Summer School, 31st Course: From Supersymmetry to the
Origin of Space-Time},
Ettore Majorana Centre, Erice, July 4-12
1993 ; hep-th/9403133, `Subnuclear Series' Vol. 31, 
(World Scientific, Singapore 1994), p.1;
\par For a pedagogical review of this approach see: D.V. Nanopoulos, 
Riv. Nuov. Cim. 17 (1994), 1.  


\bibitem{EMNW} J. Ellis, N.E. Mavromatos, D.V. Nanopoulos and
E. Winstanley, Mod. Phys. Lett. A12 (1997), 243.

\bibitem{Dbrane} J. Dai, R.G. Leigh and J. Polchinski, 
Mod. Phys. Lett. A4 (1989), 2073; 
\par J. Polchinski, Phys. Rev. D50 (1994), 6041; Phys. Rev. Lett. 
75 (1995), 184; 
\par C. Bachas, Phys. Lett. B 374 (1996), 37; 
\par J. Polchinski, S. Chaudhuri and C. Johnson, 
preprint hep-th/9602052, and references therein. 
\bibitem{counting} A. Strominger and C. Vafa, Phys. Lett. B379 (1996), 99;
\par C.G. Callan and J. Maldacena, Nucl. Phys. B472 (1996), 591;
\par A. Strominger and J. Maldacena, Phys. Rev. Lett. 77 (1996), 423.
\bibitem{EMND} J. Ellis, N.E. Mavromatos, D.V. Nanopoulos, 
Int. J. Mod. Phys. A12 (1997), 2639; hep-th/9611040, Int. J. Mod. 
Phys. A, in press. 
\bibitem{HR} S. Hawking and M. Taylor-Robinson, Phys. Rev. D55 (1997), 7680. 
\bibitem{Amati} D. Amati, preprint hep-th/9706157, and references therein.
\bibitem{recoiltime} I. Kogan, N.E. Mavromatos 
and J.F. Wheater, Phys. Lett. B387 (1996), 483. 
\bibitem{kogan} I. Kogan, Proc. {\it Particles and Fields 91} 
(eds. D. Axen, D. Bryman and M. Comyn, World Scientific, Singapore 1992), 
p.837. 
\bibitem{aben} I. Antoniadis, C. Bachas, J. Ellis and D.V. Nanopoulos,
Nucl. Phys. Phys. Lett. B211 (1988), 393; Nucl. Phys. B328 (1989), 117. 
\bibitem{mm} N.E. Mavromatos and J.L. Miramontes, 
Phys. Lett. B201 (1988), 473. 
\bibitem{birrell} N.D. Birrell and P.C.W. Davies,
{\em {Quantum Fields in Curved Space}} (Cambridge University
Press (1982)). 
\bibitem{unruh} W. Unruh, Phys. Rev. D14 (1976), 870. 
\bibitem{emnrad} J. Ellis, N.E. Mavromatos and D.V. Nanopoulos, 
Phys. Lett. B276 (1992), 56; {\it ibid.} B284 (1992), 27, 43. 
\bibitem{diamand} C. Chiou-Lahanas, G.A. Diamandis, B.C. Georgalas, 
X.N. Maintas and E. Papantonopoulos,  
Phys. Rev. D52 (1995), 5877;
\par C. Chiou-Lahanas, G.A. Diamandis, B.C. Georgalas, A. Kapella-Economou 
and 
X.N. Maintas, Phys. Rev. D54 (1996), 6226.  
\bibitem{lizzi} F. Lizzi and N.E. Mavromatos, Phys. Rev. D55 (1997), 7859.
\bibitem{fischler} W. Fischler and L. Susskind, Phys. Lett. 
B171 (1986), 383; {\it ibid.} 173 (1986), 262. 
\bibitem{aspects} J. Ellis, N.E. Mavromatos, D.V. Nanopoulos, 
Lectures presented at the conference {\it Phenomenology 
of Unification from Present to Future}, Roma, March 23-26 1994
(World Scientific, Singapore 1994), p. 187; hep-th/9405196. 
\bibitem{emninfl} J. Ellis, N.E. Mavromatos and D.V. Nanopoulos, 
Mod. Phys. Lett. A10 (1995), 1685.

\bibitem{vernon} R. Feynman and F. Vernon, 
Ann. Phys. (N.Y.) 24 (1963), 118. 


\bibitem{hu} B. Hu and A. Matacz, Phys. Rev. D51 (1995), 1577;
\par B.L. Hu and A. Matacz,
Phys. Rev. D49 (1994), 6612; 
for an application to the inflation problem see: A. Matacz, gr-qc/9604022,
Phys. Rev. D55 (1997), 1860.

\end{thebibliography}
\end{document}